\documentstyle[12pt]{article}
\title{Stochastic Quantization of the
Chern-Simons Theory}
\author{L.F.Cugliandolo\thanks{Supported by ``Commission des Communaut\'es
Europ\'eennes" (DG XII-G-3)}\\
Dipartimento di Fisica, Universit\`a di Roma, ``La Sapienza",\\
I-00185 Roma, Italy\\
G.L.Rossini\thanks{CONICET-Argentina} and
F.A.Schaposnik\thanks{Investigador CIC-Buenos Aires} \\
Departamento de F\'{\i}sica, Universidad Nacional de La Plata\\
C.C. 67, 1900 La Plata\\
Argentina}
\date{}

\begin{document}
\newcommand{\beq}{\begin{equation}}
\newcommand{\eeq}{\end{equation}}
\newcommand{\barray}{\begin{eqnarray}}
\newcommand{\earray}{\end{eqnarray}}
\renewcommand{\theequation}{\thesection .\arabic{equation}}

\maketitle
\begin{abstract}

We discuss Stochastic Quantization of $d$=3 dimensional
non-Abel\-ian Chern-Simons theory. We demonstrate that the introduction of
an appropriate regulator in the Langevin equation yields a well-defined
equilibrium limit, thus leading to the correct propagator. We also analyze
the connection between $d$=3 Chern-Simons and $d$=4 Topological Yang-Mills
theories showing the equivalence between the
corresponding regularized partition functions.
We study the construction of topological invariants and
the introduction of a non-trivial kernel as an alternative
regularization.

\end{abstract}
\newpage
\section {Introduction}

The construction of Quantum Field Theories which are independent of the
space-time metric, Topological Quantum Field Theories (TQFTs), has shown to
be a fruitful area of research, with applications both in mathematics and in
physics \cite{Wi1}-\cite{Bau}(See \cite{Bir}
for a complete list of references).
 In TQFTs there are no local excitations and then the only observables
are topological invariants. In this way, Witten has shown how to obtain
Donaldson invariants in 4-dimensional manifolds \cite{Wi1}, Jones polynomials
of knot theory \cite{Wi2}, etc. From the point of view of physics, TQFTs
have been investigated in connection with 2 and 3-dimensional gravity models,
2-dimensional conformal field theories, etc.

Quantization of Topological Field Theories can be envisaged using different
approaches. For example, the path-integral method leads
to explicit expressions for the partition function and other topological
invariants \cite{Wi1}-\cite{BRT}.
Alternatively, one can quantize in the functional Schr\"{o}dinger
representation \cite{Jac},
use canonical quantization
for systems with constraints \cite{Wi2,Hor1} or Stochastic Quantization
(SQ) \cite{Bau}. These two last  approaches have revealed, as a bonus,
interesting connections between models in different number of space-time
dimensions \cite{Yu}-\cite{CLS}.

Moreover, the SQ approach \cite{PW}
to TQFTs has clarified
the construction of supersymmetric models by exploiting the connection
between Langevin equations and Nicolai  maps which trivialize the respective
models. This connection, introduced by Parisi-Sourlas \cite{PS} and
Cecotti-Girardello \cite{CG} in low dimensions ($d<4$) was extended
to $d\geq 4$
when quantization of TQFTs was undertaken using the Batalin-Vilkovisky
approach \cite{LP,BRT}.

Many aspects related to SQ of TQFTs remain to be fully understood. It is
the purpose of the present work to address to some of them. In particular,
we carefully discuss the problem of convergence towards equilibrium
of the stochastic process associated to non-Abelian Chern-Simons theory.
To this end, we introduce an appropriate regulator in the associated
Langevin equation showing that the resulting perturbative stochastic diagrams
lead to the set of ordinary Feynman diagrams for non-Abelian Chern-Simons
theory. We also show that the connection between $d$=3 Chern-Simons model
and $d$=4 Topological Yang-Mills (TYM) theory, which was
originally established \cite{Yu}-\cite{CLS}
 without taking care of the convergence problem still holds when a
regulating scheme is taken into account.

To more precisely understand the problems to be discussed, let
us briefly review in this Introduction the basic features arising in the
quantization of TQFTs.
Firstly, since the action $S[\phi]$  for fields $\phi$ in TQFTs
does not involve at all the metric of the space-time manifold, several
problems regarding quantization are posed. Within the
path-integral approach, the metric could enter through BRST terms
needed to fix its large ({\it topological})
symmetry. One should then be sure that the topological character of the
theory is not spoiled by quantization. In fact consider the ``quantum action"
$S_Q[\phi]$ appearing in the path-integral defining the partition function:

\beq
S_Q[\Phi,g_{\mu\nu}] = S[\phi] +\{Q,W[\Phi,g_{\mu\nu}]\} .    \label{A}
\eeq
Here $\{Q,W\}$ stands for the BRST commutator of W, a functional of the
whole field content $\Phi$ ($\Phi$ includes the original fields $\phi$,
Lagrange multipliers, ghosts, ghost for ghosts, ...) and also of the metric.
It is easy to see that:

\beq
<\frac{\delta S_Q}{\delta g_{\mu\nu}}> =
<\{Q,\frac{\delta W}{\delta g_{\mu\nu}}\}>
= 0                                                 \label{B}
\eeq
due to the vanishing of BRST commutators vacuum expectation values
\cite{Wi1}.
However, in order to show that the partition function is
indeed metric independent,
one has to specify an invariant measure (being the na\"{\i}ve measure $D\Phi$
metric dependent). This can be done by taking as
integration variables not the original fields $\Phi$ but appropriate tensorial
densities \cite{Fu}-\cite{Ra1}:

\beq
\hat \Phi^{a} \equiv g^{\frac{\omega_a}{2}} \Phi^a       \label{E}
\eeq
(here $ \omega_a$ is a weight associated to $\Phi^a$ and $g = det g_{\mu\nu}$).

The partition function $Z$ for the TQFT is then defined as \cite{CLS2}:

\beq
Z = \int D\hat\Phi ~exp(-S_Q[\hat\Phi,g_{\mu\nu}])  \label{C}
\eeq
with $\hat \Phi$ and $g_{\mu\nu}$ taken as independent variables.
 One can now easily show from eqs.(\ref{B}) and (\ref{E}) that the topological
character of the model is preserved:

\beq
<T_{\mu\nu}> \equiv -\frac{2}{\sqrt{g}} \frac{\delta \log{Z}}
{\delta g_{\mu\nu}} = 0.\label{F}
\eeq

The same care regarding integration variables has to be taken in defining
the partition function for TQFTs within the stochastic quantization approach
\cite{CLS}. There are however new problems in this last approach,
associated
with TQFTs peculiarities. In particular, being the action
$S[\phi]$ metric independent, it remains unchanged when passing from Minkowski
to Euclidean space. This poses of course a problem since Stochastic
Quantization was originally formulated for real Euclidean actions. The problem
of how to handle complex actions has nevertheless been carefully analyzed
\cite{Par,Amb} and it has
recently received much attention \cite{Pol,Wu}.

In a particularly attractive TQFT, and here we come to a main point
in this paper,
there is another source of problems conspiring against convergence
towards
equilibrium within the stochastic approach. Indeed, consider 3-dimensional
Chern-Simons (CS) theory. Since the corresponding action is linear in
derivatives one should expect  non-convergence of the
associated stochastic process towards equilibrium.
In order to solve this problems, refined treatments for the Langevin equation
of a CS theory have been proposed. They basically rest
on the introduction of non-trivial kernels \cite{Wu,Fe}
(this compares to the treatment
of fermionic models) both for Abelian and non-Abelian cases
 or Maxwell terms \cite{Wu}, for the Abelian case, as regulators.
In both cases it has been proven that conventional propagators can be obtained
very simply reproducing the standard results for Chern-Simons theories.

As stated above, for {\it non-Abelian} CS theory, convergence of the stochastic
process has been only discussed by introducing an appropriate kernel \cite{Wu}.
This approach poses several difficulties when studying the connection between
$d$=3 CS theory and TYM theory. We shall then follow
an alternative regularization approach by extending the method employed
in \cite{Wu} (based on the introduction of a regulating Maxwell term for
the Abelian theory) to the non-Abelian case. We shall analyze the
convergence of the associated stochastic process towards
equilibrium and establish the connection
between $d$=3 CS theory and $d$=4 TYM theory.

The paper is organized as follows: in Section 2 we introduce a $tr F_{ij}^2$
regulating term in the Langevin equation and show, up to second order in
perturbation theory, that stochastic diagrams reproduce the standard Feynman
rules for CS theory. Then, in Section 3 we reanalyze the connection between
$d$=3 CS and $d$=4 TYM theories and study topological features of the
resulting 4-dimensional effective theory. We also discuss in this Section
the alternative regularization scheme based on the introduction of a
non-trivial kernel. Finally, in Section 4 we present a discussion of our
results.

\section{Regulating the stochastic process}
\setcounter{equation}{0}

As we stated in the Introduction, SQ of $d$=3
Chern-Simons theory faces problems of convergence towards equilibrium when
the na\"{\i}ve Langevin equation is used. These problems are rooted in two main
features of Topological Field Theories. Firstly, being the corresponding
action metric-independent, it remains unchanged when passing from Minkowski
to Euclidean space, this leading to a purely imaginary Euclidean action.
Secondly, being the CS theory linear in its derivatives,
it does not ensure a finite equilibrium limit.

Wu and Zhu \cite{Wu} have discussed how to handle both problems.
On the one hand, one can employ the so-called complex Langevin equation
approach \cite{Par}: no inconsistency then arises due to the presence of
the {\it i}-factor in front of the Euclidean CS action. Concerning
the second point, these authors introduce as a regulator
 a $F_{ij}^2$ term when studying
the Abelian theory. At least in this last case,
 one ends up with a stochastic process tending to
equilibrium and leading to standard results for propagators once the regulator
is switched off.

In this Section we shall discuss this problem of convergence for the
{\it non-Abelian} Chern-Simons theory within the SQ scheme. To this end
we introduce a $tr F_{ij}^2$ term and then analyze the convergence
of the stochastic process (in the next Section we shall discuss the
alternative method of introducing a kernel in the Langevin equation,
as proposed in Refs.\cite{Wu,Fe}).

The Euclidean action for $d$=3 non-Abelian Chern-Simons theory
with a $tr F_{ij}^2$ regulating term reads (compare with the topologically
massive Yang-Mills theory \cite{Ja}):

\beq
S = S_{CS} + S_{\Lambda} \label{2.1}
\eeq
with:

\beq
S_{CS}=- \frac{i\kappa}{8\pi} tr \int_{M_3}d^3x\epsilon_{ijk}(A_i\partial_j A_k
+ \frac{2}{3} eA_iA_jA_k) \label{2.2}
\eeq

\beq
S_{\Lambda} = \frac{1}{4\Lambda^2} tr \int_{M_3}d^3x F_{ij}^2 \label{2.3}
\eeq
Here $A_i$ ($i = 1,2,3$) takes values in the Lie algebra of the gauge
group $G$, with generators $t^a$ ($tr~t^at^b=\delta^{ab}$), $\kappa =
ke^2$ with $k\in Z$, and $M_3$ is the 3-dimensional compact manifold.
In eq.(\ref{2.3}), $\Lambda$ is
a regulator which will be set to infinity at the
end of the computations. This type of regulating Yang-Mills term has already
been used \cite{Chen} to study renormalization of the non-Abelian Chern-Simons
theory. It is important to stress that for finite $\Lambda$ the model becomes
metric-dependent. Consequently it is important to determine whether the
resulting effective action depends or not on the metric at the quantum
level (i.e. whether the quantized theory mantains diffeomorphism invariance,
a symmetry of the original Chern-Simons theory). In fact, Chen et al.
\cite{Chen} have argued that diffeomorphism anomaly is indeed absent and
our results confirm this fact.

We take as Langevin equation associated with action (\ref{2.1}) the following
one:

\beq
\partial_{t} A_i(x,t) =- \frac{\delta S}{\delta A_i}(x,t) +
\eta_i(x,t)
\label{2.4}
\eeq
where  $t$ is the stochastic time ($t \in I \equiv [0,T]$) and
$\eta_i$ is a gaussian noise taking values in the Lie algebra of the group
G. A drift term of the form $D_i\Omega$
( with $D_i$ the covariant derivative and $\Omega$ an arbitrary function)
can be introduced as a way of (stochastic) gauge fixing \cite{Z}.
 Moreover, in the analysis of the connection between $d$=3
Chern-Simons and $d$=4 TYM theories, drift terms naturally
appear when one derives Langevin equations by imposing commutativity between
BRST transformations and stochastic time evolution \cite{Bau2}.
Also, the drift term
provides in this context a natural way for introducing an $A_0$ component
for the gauge field in the route towards the effective 4-dimensional theory.
We shall come back to this point in the next Section.

In the SQ approach \cite{PW}, vacuum expectation values (v.e.v.) are computed
as the limit -for stochastic time going to infinity- of stochastic
expectation values:

\beq
\lim_{T \to \infty} <F[A_\eta]>_\eta = <F[A]>
\label{2.5}
\eeq
Here $A_\eta$ denotes the solution of Langevin equation giving A as a
functional
of the noise; the r.h.s. of eq.(\ref{2.5}) is the v.e.v. of
the functional $F[A]$
for the quantum field theory defined from action (\ref{2.1}).
Concerning stochastic expectation values $<~>_\eta$, they are computed from
the noise correlation functions:
\barray
<\eta^a_i>_\eta &=& 0 \nonumber \\
<\eta^a_i(x_1,t_1)\eta^b_j(x_2,t_2)>_\eta &=& 2\delta^{ab} \delta_{ij}
\delta^{(3)}(x_1-x_2)\delta(t_1-t_2).    \label{2.6}
\earray

Using the explicit form (\ref{2.1})-(\ref{2.3}) for $S$,
Langevin equation in momentum space becomes:
\beq
\dot{A}^a_i = -\frac{1}{\Lambda^2}k^2 P_{ij}(k)A^a_j
+\frac{\kappa}{4\pi}\epsilon_{ijl}k_lA^a_j +Y^a_i(k,t), \label{2.7}
\eeq
where $\dot{A}^a_i = \partial A^a_i /\partial t$ and  $P_{ij}(k)$
and $Y^a_i(k,t)$ are defined as:

\beq
P_{ij}(k) = \delta_{ij}-\frac{k_ik_j}{k^2}, \label{fi1}
\eeq

\barray
\lefteqn{Y^a_i(k,t)=\eta^a_i(k,t)+}   \\
& &+\frac{e}{(2\pi)^{3/2}} \int d^3p~d^3q~ \delta(k-p-q)
V^{abc}_{ijk}(k,-p,-q) A^b_j(p,t)A^c_k(q,t) + \nonumber \\
& &+\frac{e^2}{(2\pi)^3} \int d^3p~d^3q~d^3r \delta (k-p-q-r) W^{abcd}_{ijkl}
A^b_j(p,t)A^c_k(q,t)A^d_l(r,t) .    \nonumber \label{2.8}
\earray
In the last equation $V^{abc}_{ijk}$ and $W^{abcd}_{ijkl}$ are, respectively,
the three-point and four-point vertex factors (listed in Table~1 together
with the corresponding diagrams). Notice that the three-point vertex factor
is just $\frac{1}{2}$ of the corresponding one in the conventional Feynman
rule,
while the four-point vertex factor is $\frac{1}{6}$ of the corresponding
conventional one.

\begin{table}
\begin{tabular}{c|c|c} \hline \hline
DIAGRAM & NOTATION & FORMULA \\ \hline
\makebox[3cm]{} & & \\
 & $eV_{ijk}^{abc}(k,k_1,k_2)$ & $-\frac{i}{2}e f^{abc}
[\frac{\kappa}{4\pi}\epsilon_{ijk}+\frac{1}{\Lambda^2}((k-k_1)_k\delta_{ij}+$
\\
 & & $+(k_1-k_2)_i\delta_{jk}
+(k_2-k)_j\delta_{ik})]$ \\
 & &  \\ \hline
 & & \\
 &  & $-\frac{1}{6}\frac{e^2}{\Lambda^2}[f^{abe}f^{cde}
(\delta_{ik}\delta_{jl}-\delta_{il}\delta{jk}) +$ \\
& $e^2W_{ijkl}^{abcd}$ & $+f^{ace}f^{bde}(\delta_{ij}\delta_{kl}-
\delta_{il}\delta{jk})+$ \\
& & $+f^{ade}f^{cbe}(\delta_{ik}\delta_{jl}-\delta_{ij}\delta{kl})]$
\\
 & & \\ \hline
\end{tabular}
\caption{Diagrams for stochastic vertices.}
\end{table}

The Langevin equation can be solved perturbatively, and for this purpose
it is convenient to write it as an integral equation. Using $A^a_i(k,0)=0$
as initial condition, one readily obtains
\beq
A^a_i(k,t)=\int_0^\infty dt'\; G^{ab}_{ij}(k,t-t')Y^b_j(k,t') ,
\label{2.9}
\eeq
where $G^{ab}_{ij}(k,t-t')$ is the zeroth order (causal) Green function
associated with eq.(\ref{2.7}):
\beq
\delta^{ab}(\delta_{ij}\frac{\partial}{\partial t}+\frac{k^2}{\Lambda^2}
P_{ij}(k)-\frac{\kappa}{4\pi} \epsilon_{ijl}k_l)
G^{bc}_{jm}(k,t-t')= \delta^{ac}\delta_{im}\delta(t-t'), \label{2.10}
\eeq
which yields
\barray
G^{ab}_{ij}(k,t-t')=\theta(t-t') \delta^{ab}\left\{
[P_{ij}(k)cos(\frac{\kappa}{4\pi}k(t-t'))+\right. \nonumber \\
\left. +\frac{\epsilon_{ijl}k_l}{k} sin(\frac{\kappa}{4\pi}k(t-t'))]
e^{-\frac{1}{\Lambda^2}k(t-t')} + \frac{k_ik_j}{k^2}\right\} . \label{2.11}
\earray

It is convenient to write eq.(\ref{2.9}) in a symbolic way as
\beq
A=G(\eta+eVAA+e^2WAAA).  \label{2.12}
\eeq
The perturbative expansion of the solution then reads:
\barray
A &=& G\eta +eGV(G\eta)(G\eta) +e^2W(G\eta)(G\eta)(G\eta)+ \label{2.13}\\
& & +e^2GV(GV(G\eta)(G\eta))(G\eta)+e^2GV(G\eta)(GV(G\eta)(G\eta))+\cdots
\nonumber
\earray
which is graphically represented in Figure~1.

\begin{figure}
\vspace{2cm}
\begin{tabbing}
espacio--- \=   \kill
\> {\it A}=
\end{tabbing}
\caption{Perturbative expansion of $A_i^a(k,t)$.}
\end{figure}

In order to compute an $n$-point correlation function
$<A^{a_1}_{i_1}(k_1,t)\cdots \linebreak A^{a_n}_{i_n}(k_n,t)>_{\eta}$,
one proceeds as follows: for each $A^a_i(k,t)$
one draws a diagrammatical expansion as that shown in Figure 1, then
one takes the
average over the noise by joining all crosses (representing the noise
sources), pair-wise, in all possible ways and finally one sums
up the resulting
"stochastic diagrams", as implied by the noise correlation
functions (see eq.(\ref{2.6})).

Notice that every diagram is built up from two kinds of propagators:

\noindent a) the un-crossed propagator given by the stochastic Green function
(\ref{2.11}):
\barray
G^{ab}_{ij}(k,t-t')=\theta(t-t') \delta^{ab}
\left\{ [P_{ij}(k) cos(mk(t-t'))+\right. \nonumber \\
\left. +\frac{\epsilon_{ijl}k_l}{k} sin(mk(t-t'))]
e^{-\frac{1}{\Lambda^2}k(t-t')} + \frac{k_ik_j}{k^2}\right\} .
\earray
b) the crossed propagator which corresponds to the lowest order term in
$<A^a_i(k,t)A^b_j(-k,t')>_{\eta}$:
\barray
\lefteqn{D^{ab}_{ij}(k;t,t')= \delta^{ab} 2t_{min} \frac{k_ik_j}{k^2} + }
\nonumber \\
& &+\delta^{ab}\frac{1}{\frac{k^2}{\Lambda^4}+m^2}
\left\{ e^{-\frac{k^2}{\Lambda^2}|t-t'|} \left[ (\frac{1}{\Lambda^2}
P_{ij}(k) + \frac{m}{k}\frac{\epsilon_{ijl}k_l}{k})
cos(mk|t-t'|)+\right. \right. \nonumber \\
& &\left. +(-\frac{m}{k}P_{ij}(k)+\frac{1}{\Lambda^2}
\frac{\epsilon_{ijl}k_l}{k})sin(mk|t-t'|) \right] + \nonumber \\
& &-e^{-\frac{k^2}{\Lambda^2}(t+t')} \left[ (\frac{1}{\Lambda^2}
P_{ij}(k) + \frac{m}{k}\frac{\epsilon_{ijl}k_l}{k})
cos(mk(t+t'))+\right. \nonumber \\
& &+\left.\left.(-\frac{m}{k}P_{ij}(k) +\frac{1}{\Lambda^2}
\frac{\epsilon_{ijl}k_l}{k})sin(mk(t+t')) \right]\right\} \label{2.15}
\earray
In the last equation $m$ stands for $\kappa / 4\pi$, usually
called the "topological mass" of the model \cite{Ja},
and $t_{min}$ is the minimum
between $t$ and $t'$. From the above expressions one
recognizes a linearly divergent longitudinal
part which is common in the SQ of all gauge theories without gauge fixing.
This term can be handled by stochastic gauge fixing \cite{Z} (see below).

If not for the presence of two kinds of propagators, each stochastic diagram
has the form of an ordinary Feynman diagram. Though the rules for these
stochastic propagators are different from the Feynman rules, it has been
proven for several models that the sum of all stochastic diagrams with
the same topology yields exactly the usual result for the corresponding
Feynman diagram.
In particular, Namiki et al. \cite{Na} have shown that, up to second order,
this is the case
for pure Yang-Mills theory. In the same spirit we shall
study the second order corrections to the propagator of the non-Abelian
Chern-Simons model (with the $tr F_{ij}^2$ regulator term) and we shall
show that
its transverse part reproduces the correct field-theoretical propagator.

As stressed above, correlation functions
$<A_i^a (k_1,t)A_j^b (k_2,t')>$
contain terms which diverge as $t$ and $t'$ go to infinity. These divergent
terms are characteristic of gauge non-invariant quantities and will cancel
out if one constructs gauge-invariant objects like $<trF_{ij}^2>$. Consequently
we shall only keep in what follows those terms which remain finite as $t$
or $t'$ go to infinity. Using (\ref{2.13}) and averaging over $\eta$ we
then get for the
correlation function, up to second order:

\beq
<A_i^a(k,t)A_j^b(k',t)> = \delta(k+k'){\cal D}_{ij}^{ab}(k,t) \label{2.16}
\eeq
with:
\beq
{\cal D}_{ij}^{ab}(k,t) = (a) + 2(b) + 2(c) + 2(d) + 3(c) + (f), \label{2.17}
\eeq
where $(a), (b), ...,(f)$ stand for the contributions from diagrams shown
in Figure~2. As an example, diagram $(b)$ represents the following
contribution:

\barray
\lefteqn{2(b) = \frac{2e^2}{(2\pi)^3} \int d^3k_1~d^3k_2 \delta (k-k_1-k_2)
V^{cde}_{klm}(k,-k_1,-k_2) V^{fgh}_{qnp}(-k,k_1,k_2) \times} \nonumber \\
& &\times \int_0^{\infty}dt'\int_0^{\infty}dt''G_{ik}^{ac}(k,t-t')
D_{ln}^{dg}(k_1;t',t'') D_{mp}^{eh}(k_2;t',t'')
G_{jq}^{bf}(-k,t-t''). \nonumber\\
& &\label{2.18}
\earray
Here it should be noted that longitudinal components of external lines can
be discarded either because they vanish or because they cancel out when
computing gauge invariant or gauge covariant contributions. This is carefully
shown in Ref.\cite{Na} for the pure Yang-Mills model and the demonstration
trivially
applies to the present model. Concerning longitudinal components appearing
in internal lines, it has been conjectured by Parisi and Wu \cite{PW} that
they arrange themselves to give just those contributions which are
conventionally associated with Faddeev-Popov ghost effects. Although a general
proof for this conjecture is lacking, it has been explicitely confirmed
in several examples \cite{DH}. In particular, Namiki et al. \cite{Na} have
proved that, up to second order, in the case of pure Yang-Mills theory
internal lines do arrange to reproduce the ghost contributions. Since
the longitudinal part of the propagators of this last theory coincide with
those arising in the CS model (eqs.(\ref{2.15}) and (\ref{2.16})),
we conclude that also in the present case ghost effects are accounted by
longitudinal parts of internal lines and hence we shall only consider in
diagrams $(a),(b),...,(f)$ transverse internal lines.

\begin{figure}
\vspace{10cm}
\caption{Graphical representation of ${\cal D}_{ij}^{ab}(k,t)$.}
\end{figure}

Computation of contributions from different diagrams is standard but tedious.
For the sake of brevity we shall not present each one separately.
Just as an example,
the contribution from diagram $(b)$ takes the form:

\barray
2(b)=  \frac{e^2}{(2\pi)^3}&\int& d^3k_1~d^3k_2 \delta (k-k_1-k_2)
V^{ade}_{klm}(k,-k_1,-k_2) V^{bde}_{qnp}(-k,k_1,k_2) \nonumber \\
&\times& I_{iklnmpqj}(k,k_1,k_2,m,\Lambda ), \label{2.19}
\earray
where
\barray
\lefteqn{I_{iklnmpqj}(k,k_1,k_2,m,\Lambda )=} \nonumber \\
& &=P_{ik}(k) P_{ln}(k_1) P_{mp}(k_2) P_{jq}(k) F^{(1)}
+P_{ik}(k) P_{ln}(k_1) P_{mp}(k_2) \frac{\epsilon_{jqv}k_v}{k}
F^{(2)}+ \nonumber\\
& &+P_{ik}(k) P_{ln}(k_1) \frac{\epsilon_{mpu}k_{2u}}{k_2} P_{jq}(k)
F^{(3)}
+P_{ik}(k) P_{ln}(k_1) \frac{\epsilon_{mpu}k_{2u}}{k_2}
\frac{\epsilon_{jqv}k_v}{k} F^{(4)} + \nonumber\\
& &+P_{ik}(k) \frac{\epsilon_{lns}k_{1s}}{k_1} P_{mp}(k_2)
P_{jq}(k) F^{(5)}
+P_{ik}(k) \frac{\epsilon_{lns}k_{1s}}{k_1} P_{mp}(k_2)
\frac{\epsilon_{jqv}k_v}{k} F^{(6)} +  \nonumber\\
& &+P_{ik}(k) \frac{\epsilon_{lns}k_{1s}}{k_1} \frac{\epsilon_{mpu}k_{2u}}{k_2}
P_{jq}(k) F^{(7)}
+P_{ik}(k) \frac{\epsilon_{lns}k_{1s}}{k_1} \frac{\epsilon_{mpu}k_{2u}}{k_2}
\frac{\epsilon_{jqv}k_v}{k} F^{(8)}+ \nonumber\\
& &+\frac{\epsilon_{ikr}k_r}{k} P_{ln}(k_1) P_{mp}(k_2) P_{jq}(k)
F^{(9)}
+\frac{\epsilon_{ikr}k_r}{k} P_{ln}(k_1) P_{mp}(k_2)
\frac{\epsilon_{jqv}k_v}{k} F^{(10)}+ \nonumber\\
& &+\frac{\epsilon_{ikr}k_r}{k} P_{ln}(k_1) \frac{\epsilon_{mpu}k_{2u}}{k_2}
P_{jq}(k) F^{(11)}
+\frac{\epsilon_{ikr}k_r}{k} P_{ln}(k_1) \frac{\epsilon_{mpu}k_{2u}}{k_2}
\frac{\epsilon_{jqv}k_v}{k} F^{(12)} +\nonumber\\
& &+\frac{\epsilon_{ikr}k_r}{k} \frac{\epsilon_{lns}k_{1s}}{k_1} P_{mp}(k_2)
P_{jq}(k) F^{(13)}+
\frac{\epsilon_{ikr}k_r}{k} \frac{\epsilon_{lns}k_{1s}}{k_1}
P_{mp}(k_2) \frac{\epsilon_{jqv}k_v}{k} F^{(14)}+ \nonumber\\
& &+ \frac{\epsilon_{ikr}k_r}{k} \frac{\epsilon_{lns}k_{1s}}{k_1}
\frac{\epsilon_{mpu}k_{2u}}{k_2} P_{jq}(k) F^{(15)}+
\frac{\epsilon_{ikr}k_r}{k}~ \frac{\epsilon_{lns}k_{1s}}{k_1}~
\frac{\epsilon_{mpu}k_{2u}}{k_2}~ \frac{\epsilon_{jqv}k_v}{k} F^{(16)}
\nonumber\\
{}~ \label{2.20}
\earray
Here $F^{(i)} = F^{(i)}(k,k_1,k_2,m,\Lambda)$
are ratios of polynomials in momenta which arise after
integrating the fictitious time in the vertices and taking the limit $t
\to \infty$. Contributions from the rest of the diagrams having the same
topology as $(b)$ have analogous form. Computations were carried out using
REDUCE and the answer is:

\barray
\lefteqn{2(b) + 2(c) + 2(d) =} \nonumber \\
& &2\frac{e^2}{(2\pi)^3} \int d^3k_1~d^3k_2 \delta (k-k_1-k_2)
V^{ade}_{klm}(k,-k_1,-k_2) V^{fgh}_{bde}(-k,k_1,k_2) \times \nonumber \\
& &\frac{(\frac{1}{\Lambda^2}P_{ik}(k)+\frac{m}{k^2} \epsilon_{ikr}k_r)}
{\frac{k^2}{\Lambda^4}+m^2}
\frac{(\frac{1}{\Lambda^2}P_{ln}(k_1)+\frac{m}{k_1^2} \epsilon_{lns}k_{1s}
)}{\frac{k_1^2}{\Lambda^4}+m^2} \nonumber \\
& &\frac{(\frac{1}{\Lambda^2}P_{mp}(k_2)+\frac{m}{k_2^2}
 \epsilon_{mpu}k_{2u})}{\frac{k_2^2}{\Lambda^4}+m^2}
\frac{(\frac{1}{\Lambda^2}P_{jq}(k)+\frac{m}{k^2} \epsilon_{jqv}k_v)}
{\frac{k^2}{\Lambda^4}+m^2}\label{2.21}
\earray
An analogous calculation for diagrams $(e)$ gives:

\barray
\lefteqn{3(e)= 3\frac{e^2}{(2\pi)^3}\int d^3k_1~W^{addb}_{klmn}
\times} \label{2.22} \\
& &\frac{(\frac{1}{\Lambda^2}P_{ik}(k)+\frac{m}{k^2} \epsilon_{ikr}k_r)}
{\frac{k^2}{\Lambda^4}+m^2}
\frac{(\frac{1}{\Lambda^2}P_{lm}(k_1)+\frac{m}{k_1^2}
\epsilon_{lmu}k_u)} {\frac{k_1^2}{\Lambda^4}+m^2}
\frac{(\frac{1}{\Lambda^2}P_{jn}(k)+\frac{m}{k^2} \epsilon_{jns}k_s)}
{\frac{k^2}{\Lambda^4}+m^2}   \nonumber
\earray

Taking into account the relations between stochastic and conventional Feynman
vertices,
it is now evident that eqs.(\ref{2.21}) and (\ref{2.22}) coincide,
respectively, with the
gluon loop and the tadpole terms in the conventional Landau gauge field
theory (see for instance \cite{PR}).
Finally, diagram $(f)$ vanishes due to antisymmetry of the three-point vertex.
Hence we arrive to:

\vspace{1cm}
\noindent  ({\it a})+2({\it b})+2({\it c})+2({\it d})$=$
\beq
    \label{2.24}
\eeq
\vspace{2cm}

\noindent 3({\it e})+({\it f})$=$
\beq
    \label{2.25'}
\eeq

Here wavy lines represent the standard (bare) propagator for topologically
massive gauge theory, i.e. $d$=3 Yang-Mills theory plus a CS term (compare
e.g. Ref.\cite{PR}):
\beq
\makebox[3cm]{}
=\frac{(\frac{1}{\Lambda^2}P_{ij}(k)+\frac{m}{k^2} \epsilon_{ijr}k_r)}
{\frac{k^2}{\Lambda^4}+m^2}  \label{2.25}
\eeq
Thus, the r.h.s. in (\ref{2.24}) and (\ref{2.25'}) reproduce the standard
second order result in perturbation theory.
Concerning the resulting three-point and four-point vertices,
they coincide with the standard vertices for topologically massive gauge
theory and they can be read off from
those presented
in Table 1 (provided we disregard the factors $\frac{1}{2}$ and
$\frac{1}{6}$ respectively,
appearing in the stochastic scheme).

We can therefore conclude that, up to second order in perturbation,
the Stochastic Quantization scheme for non-Abelian CS model endowed
with a regulator term of the form (\ref{2.3}) leads to the correct gauge field
propagator for topologically massive Yang-Mills theory.
That is, the introduction of the $\frac{1}{\Lambda^2}tr F_{ij}^{2}$ term has
lead to a stochastic process which converges towards equilibrium. As we have
already stated, diagrams with ghosts lines should be obtained by including
the longitudinal parts of $G^{ab}_{ij}$ and $D^{ab}_{ij}$ in the internal
lines of each diagram.

If one now takes the limit of $\Lambda^2 \to \infty$ to switch off the
regulator so as to recover the pure CS gauge theory, one can easily
verify that (\ref{2.24})-(\ref{2.25}) agree with the standard (bare) expresions
(compare e.g. Ref.\cite{AG}).

We have then succeded in obtaining a convergent stochastic process for the
non-Abelian CS theory. We can then re-analize the connection between the
(regulated) stochastic partition function for CS theory and the BRST partition
function for TYM theory. This will be presented in the
next Section.

\section{The connection between CS and TYM theories}
\setcounter{equation}{0}

The connection between $d$=3 non-Abelian CS and $d$=4 TYM theories
\linebreak has been
established by Yu \cite{Yu} and Baulieu \cite{Bau2} using SQ.
Alternatively, Horowitz \cite{Hor1} has discussed the same connection
using canonical quantization for systems with constraints.

Concerning the SQ derivation, it consists in proving the equivalence between
the stochastic partition function for CS theory in the limit $T \to \infty$
and the BRST partition function for TYM theory (stochastic time $t$
provides the extra coordinate necessary to pass from $M_3$ to a $d$=4 manifold
$M_4$).

The analysis in \cite{Yu,Bau2} did not take into account the necessity of
working with an invariant measure for the path-integral which defines the
partition functions. As mentioned in the Introduction this poses a problem
since the na\"{\i}ve measure is in fact metric dependent, and hence the
topological
character of the quantum theory cannot be simply established. In
\cite{CLS} this problem was studied using the so-called Fujikawa variables.
This variables allow to define an invariant measure at the price of working
not with the original fields but with appropriate tensorial densities (see
eq.(\ref{E}) and (\ref{C})).
The outcome is that the connection between CS and
TYM partition functions is still valid when the correct integration
variables are considered.

Another important point that was not taken into account in the derivations
given by \cite{Yu}-\cite{CLS} based in SQ was that related to the
non-convergence
of the underlying stochastic process. We have proven in the previous Section
that the introduction of a regulator in the Langevin equation
yields the correct equilibrium limit
for the stochastic process and gives the standard propagators for
the non-Abelian CS theory. Once this is achieved, one has to re-examine
the question of whether the connection between CS and TYM theories
is still valid when the theory is
regulated. We address to this point in the first part of this Section and
show
that the answer is affirmative. Then we also analyze the same problem when
a non-trivial kernel (instead of a regulator) is introduced in the Langevin
equation, using a slightly different approach.

In order to proceed to the proof of the connection \`{a} la Yu-Baulieu,
we have to modify Langevin equation (\ref{2.4}) for the CS model
adding a longitudinal drift term. In fact
when one derives Langevin equations by imposing commutativity between BRST
transformations and stochastic time evolution, a drift
term naturally appears \cite{Bau2}.
This term can be in fact identified with the one
which is usually introduced in order to handle the non-convergence of gauge
dependent v.e.v.'s within the SQ approach. In the present context,
it provides a natural way for introducing an $A_0$ component for the CS
gauge field in the route towards the 4-dimensional theory. Indeed, if instead
of eq.(\ref{2.4}) we write a Langevin equation with drift term:
\beq
\partial_t A_i =-\frac{\delta S}{\delta A_i} - D_i\Omega +\eta_i,
\label{3.1}
\eeq
being $\Omega$ an arbitrary function, we can take:
\beq
\Omega(x,t)= -\frac{\kappa}{4\pi}A_0(x,t). \label{3.2}
\eeq
Here $A_0$ will be considered, as announced, the zeroth component of
the gauge field $A_{\mu} \equiv (A_i,A_0)$, now defined over a four manifold
$M_4 = M_3 \times I$. We then impose a Langevin equation for $A_0$:
\beq
\frac{\partial A_0(x,t)}{\partial t} = -\frac{\kappa}{4\pi} \partial_i A_i +
\eta (x,t)
\label{3.3}
\eeq
with $\eta$ a scalar gaussian noise taking values in the Lie algebra of
the group G.

In order to write equations (\ref{3.1}) and (\ref{3.3})
in a more appropriate way, we introduce $\partial_0=\frac{4\pi}{\kappa}
\partial_t$. We also redefine $e A_{\mu} \to A_{\mu}$ and
set $e^2=\frac{\kappa}{4\pi}$.
Langevin equations take then the form:
\beq
F_{0i}^{+}-\frac{4\pi}{\kappa\Lambda^2}D_j F_{ji} = \sqrt{\frac{4\pi}{\kappa}}
\eta_i
\label{3.4}
\eeq
and
\beq
\partial_{\mu}A_{\mu}=\sqrt{\frac{4\pi}{\kappa}} \eta. \label{3.5}
\eeq
Here
\beq
F_{\mu\nu}=\partial_{\mu}A_{\nu}-\partial_{\nu}A_{\mu}+ie[A_{\mu},A_{\nu}]
\label{3.6}
\eeq
and
\beq
F_{\mu\nu}^{+} \equiv F_{\mu\nu} + \frac{1}{2} \epsilon_{\mu\nu\alpha\beta}
F_{\alpha\beta}. \label{3.7}
\eeq

We are now ready to establish the connection between
$d$=4 TYM and $d$=3 CS theory
(regulated with the $tr F_{ij}^2$ term) when the regulator is \linebreak
switched off.
We start from the stochastic generating functional for the theory associated
with equations (\ref{3.4}) and (\ref{3.5}):
\beq
Z_{Stoch}^{CS}[\Lambda]=\int D\eta_i D\eta exp[-\frac{2\pi}{\kappa}tr
\int_{M_4} d^4x
(\eta_i^2+\eta^2)],
\label{3.8}
\eeq
where $d^4x=d^3x dx_0$
We have used in eq.(\ref{3.8}) a stochastic gaussian measure defined in
terms of the original gaussian noises $\eta_i$ and $\eta$ and not in terms
of the associated Fujikawa variables \cite{Fu}

\beq
\begin{array}{lll}
\hat{\eta_i}= g^{1/12}\eta_i & & \hat{\eta}=g^{1/4}\eta. \label{3.9}
\end{array}
\eeq

\noindent As we stated above, these last ones are
the correct variables defining an
invariant measure. However, as it was proven in \cite{CLS}, the derivation
of the connection between CS and TYM theories follows exactly the same
steps using ordinary or Fujikawa variables
and hence, for the sake of simplicity, we shall work with the original
variables although our proof can be straightforwardly presented in terms
of Fujikawa variables.

We now write $Z_{Stoch}^{CS}[\Lambda]$ as a path integral
over gauge fields $A_{\mu}$ by using
\beq
D\eta D\eta_i=J D\! A_{\mu} \label{3.10}
\eeq
with the Jacobian given by
\beq
J=det M_{i\nu} \label{3.11}
\eeq
and
\beq
M_{i\nu}= \left(
\begin{array}{c}
\frac{\delta \eta}{\delta A_{\nu}} \\  \\
\frac{\delta \eta_i}{\delta A_{\nu}}
\end{array}
\right)  \label{3.12}
\eeq
Ghost fields $(b,\chi_i)$ and $\psi_{\nu}$ (with ghost
numbers (-1,-1) and 1 respectively) can be introduced
in order to exponentiate $M_{i\nu}$:
\beq
J=\int Db D\chi_i D\psi_{\mu} exp[tr \int_{M_4} d^4x
(b,\chi_i) M_{i\nu} \psi_{\nu}] \label{3.13}
\eeq
Of course, the change of variables holds for $J\neq 0$ (which corresponds,
in the limit $\Lambda\to\infty$, to $dim {\cal M}=0$, being ${\cal M}$
the instanton moduli space associated with
$F_{\mu\nu}^+ =0$ equation ). The case $J=0$ ($dim
{\cal M}\neq 0$) can be easily treated by extending the procedure which
we describe below (see Ref.\cite{Yu} for details).

In order to end up with the complete set of ghosts characterizing BRST
invariance of TYM theory we introduce,
following \cite{Bau3}, $c$-ghosts (associated
with ordinary gauge symmetry in TYM) through the equation:
\beq
D_{\mu}D_{\mu}c-D_{\mu}\psi_{\mu}=\beta \label{3.14}
\eeq
with $\beta$ a gaussian noise. Inserting the identity
\beq
1=\int D\rho D\beta exp[-tr\int \rho\beta d^3x dt]  \label{3.15}
\eeq
in $Z^{CS}_{Stoch}$ and changing from $\beta$ to $c$
variables via eq.(\ref{3.14}) we finally have
\beq
Z^{CS}_{Stoch}=\int DA_{\mu} Db D\chi_{\alpha\beta}
D\psi_{\mu} Dc D\rho D\lambda
D\phi e^{-S_{eff}}. \label{3.16}
\eeq
Here $\lambda$ and $\phi$ are (Grassman even) ghosts (with ghost number
$-2$ and $2$ respectively), introduced in order to exponentiate the determinant
associated with the change from $\beta$ to $c$ variables. Concerning
$\chi_{\alpha\beta}$, we have traded  the three components of ghost field
$\chi_i$ for those of a self-dual antisymmetric ghost
%$\chi_{\alpha\beta}$
(again with ghost number $-1$). This is done to make contact with the
usual notation adopted in the literature \cite{Wi1,Yu,Bau2}, and, moreover,
to take into account the correct tensorial character of this field \cite{CLS}
The effective action $S_{eff}$ reads:
\barray
S_{eff}=\int_{M_4} &tr&\{ \frac{1}{2}(F_{0i}^+ -
\frac{4\pi}{\kappa\Lambda^2}
D_jF_{ji})^2 +\frac{1}{2}(\partial_{\mu}A_{\mu})^2+\rho(D_{\mu}D_{\mu}c
- D_{\mu}\psi_{\mu})+  \nonumber\\
&+&\frac{4\pi}{\kappa\Lambda^2}\chi_{0i}(\delta_{ij} D_kD_k-D_iD_j)\psi_j-
\chi_{\mu\nu}D_{[\mu}\psi_{\nu ]}-b\partial_{\mu}\psi_{\mu}+ \nonumber\\
&+&\lambda D_{\mu}D_{\mu}\phi+\lambda([\psi_{\mu},\psi_{\mu}+D_{\mu}c]+
D_{\mu}[\psi_{\mu},c]) \} d^4x. \label{3.17}
\earray
In order to write $S_{eff}$ in a more tractable way, let us introduce
Lagrange multipliers $\eta_{\alpha\beta}$ (self-dual and antisymmetric)
and $\eta$, and the following (BRST) transformations:
\beq
\begin{array}{lll}
\{Q,A_{\mu}\}=\psi_{\mu}  & & \{Q,\psi_{\mu}\}=0 \\
\{Q,b\}=\eta  & & \{Q,\eta\}=0 \\
\{Q,\chi_{\alpha\beta}\}=\eta_{\alpha\beta}  & & \{Q,\eta_{\alpha\beta}\}=0 \\
\{Q,c\}=\phi  & & \{Q,\phi\}=0 \\
\{Q,\lambda\}=\rho  & & \{Q,\rho\}=0 \\
\end{array}  \label{3.18}
\eeq
Then, $S_{eff}$ can be written in the form:
\beq
S_{eff}=\{Q,V_{\Lambda}\} \label{3.19}
\eeq
with
\barray
V_{\Lambda}=&\int_{M_4}& d^4x\; tr\{\frac{1}{4}\chi_{\alpha\beta}
(F^+_{\alpha\beta}-\frac{1}{2}\eta_{\alpha\beta})- \chi_{0i}
\frac{4\pi}{\kappa\Lambda^2}D_jF_{ji}+ \nonumber\\
&+& b(\partial_{\mu}A_{\mu} - \frac{1}{2}\eta)
-\lambda(D_{\mu}\psi_{\mu}+
D_{\mu}D_{\mu}c)\} \label{3.20}
\earray

In this way, we can easily see that
the $\Lambda$-independent part of
$S_{eff}$ coincides with the action constructed by Witten
for defining TYM theory \cite{Wi1}.
Accordingly, the BRST transformations (\ref{3.18}) correspond to
those introduced by Labastida and Pernici \cite{LP} in order to obtain TYM
as the result of gauge fixing a gaussian trivial action.
To see this correspondence
explicitly, one has to proceed to the following change of variables
\cite{OSB} :
\barray
\psi_{\mu}&\to &\psi_{\mu}+D_{\mu}c \nonumber\\
\phi &\to & \phi + \frac{1}{2}[c,c]. \label{3.21}
\earray
Then, transformations (\ref{3.18}) coincide with the BRST transformations
defined in Ref.\cite{LP} in order to handle
the large topological symmetry characteristic of TYM theory.

We can now write the following equality which turns to be one of the
main results in our work:
\beq
Z^{CS}_{Stoch}[\Lambda]=Z^{TYM}_{BRST}[\Lambda] \label{3.22}
\eeq
where
\beq
Z^{TYM}_{BRST}[\Lambda] =\int D\Phi\; e^{\{Q,V_{\Lambda}\}} \label{prima}
\eeq
and $D\Phi$ represents the integral measure over the whole field content.
Hence, the stochastic partition function for the $d$=3 CS model regulated
with a Yang-Mills term coincides with the BRST partition function for a
$d$=4 TYM model with the addition of a $\Lambda$-dependent functional
(eq.(\ref{3.20})).

The fact that the $\Lambda$-dependent terms in $S_{eff}$ appear through
$Q$-exact terms allows us to proceed one step further. Indeed, since
\beq
\frac{\delta Z^{TYM}_{BRST}[\Lambda]}{\delta\Lambda}=
-\int D\Phi \{ Q,\frac{\partial V_{\Lambda}}{\partial\Lambda} \}
e^{\{Q,V_{\Lambda}\}},
\label{3.23}
\eeq
then
\beq
\frac{1}{Z}\frac{\delta Z^{TYM}_{BRST}]\Lambda]}{\delta\Lambda}=
-<\{ Q,\frac{\partial V_{\Lambda}}{\partial\Lambda} \}>=0 \label{3.25}
\eeq
due to the vanishing of v.e.v.'s of BRST commutators \cite{Wi1}. Then, at the
partition function level we infer that, being $Z^{TYM}_{BRST}[\Lambda]$
$\Lambda$-independent, it can be evaluated in the limit $\Lambda$ going
to infinity. Hence, we can write instead of (\ref{3.22})
\beq
Z^{CS}_{Stoch}=Z^{TYM}_{BRST}, \label{3.26}
\eeq
where in the r.h.s one has now the partition function for TYM theory with
the action constructed by Witten. This does not mean that from the stochastic
point of view the regulator is not necessary. As explained in the previous
Section, in order to compute CS propagators, etc., one has to mantain the
regulator till the end of the computations. Eq.(\ref{3.26})
implies that one can use a regulator $\Lambda$ to render all quantities
appearing in the derivation of (\ref{3.22}) well-defined; at the end
of the computations, being the resulting partition
function $\Lambda$-independent, one can make $\Lambda \to \infty$ thus
confirming the connection between CS and TYM models.
In this respect, one can think of identity (\ref{3.26}) as follows: from the
CS point of view one has to mantain the regulator in order to have a convergent
stochastic process, yielding the correct stochastic averages. One then makes
$\Lambda\to\infty$ at the end of the computations. From the TYM
point of view, due to (\ref{3.25}) one can make $\Lambda \to \infty$ at
any step of the computations, not only for $Z^{TYM}_{BRST}$ but for any v.e.v.
to be computed.

Indeed, the basic formula (\ref{3.19}) also ensures that
\beq
<T_{\mu\nu}>= -\frac{2}{\sqrt{g}}
\frac{\delta\log{Z_{BRST}^{TYM}[\Lambda ]}}{\delta g^{\mu\nu}}=0.
\label{3.27}
\eeq
That is, $Z_{BRST}^{TYM}$ defines a topological model for {\it any} value
of $\Lambda$. As we discussed above, in order to prove (\ref{3.27}) one
should use Fujikawa variables but, when written in these terms,
%proceeding as in Ref.\cite{CLS},
one can still prove that $S_{eff}$ is a $Q$-exact form.
We skip details since they parallel those presented in
Ref.\cite{CLS}.

Analogously, one has, for arbitrary $\Lambda$
\beq
\frac{\partial Z_{BRST}^{TYM}[\Lambda ]}{\partial e^2}=0.
\label{3.28}
\eeq
The independence of the partition function for TQFT's on the coupling constant
has as an important consequence that one can compute it in the weak-coupling
(semi-classical) limit.
On the one hand, for TYM theory (i.e. once the limit $\Lambda\to\infty$
is taken)
one can therefore write $Z_{BRST}^{TYM}$
as a sum over contributions from the neighbourhoods of the isolated instantons
since classical minima of the TYM action correspond to solutions of
\beq
F^+_{\mu\nu}=0. \label{3.29}
\eeq
In expanding around isolated instantons it is then enough to keep only
quadratic terms and the resulting gaussian integrals give:
\beq
\frac{P\! f\! a\! f\! f\; D_F}{\sqrt{det\; \Delta_B}}, \label{3.30}
\eeq
where $P\! f\! a\! f\! f\; D_F$ denotes the Pfaffian of the operator
associated to
Fermi fields and $\Delta_B$ that associated to Bose fields in the quadratic
part of the action. Up to a sign, this ratio is 1 and then
\beq
Z_{BRST}^{TYM}=\sum_{instantons} (\pm 1), \label{3.31}
\eeq
that is, it coincides with the first Donaldson invariant.

On the other hand,
for $Z_{BRST}^{TYM}[\Lambda]$ at finite $\Lambda$, classical minima for
action $S_{eff}$ are not any more solutions of (\ref{3.29}), although in
the weak-coupling limit gaussian integrals give a ratio like (\ref{3.30})
which is still $\pm 1$. However, the partition function will yield
the first Donaldson invariant provided one takes the $\Lambda \to \infty$
limit.

It is important to notice that the $\Lambda$-independence of $Z^{TYM}_{BRST}$
extends to any (topological invariant) observable. Indeed,
the sufficient conditions for an observable $\Theta$
to give a topologically invariant
v.e.v. are \cite{Wi1}:

\noindent 1) $ \delta \Theta / \delta g^{\mu\nu}=\{Q,R\} $

\noindent 2) $ \{Q,\Theta \}=0$ (modulo those of the form $\Theta=\{Q,B\}$).

\noindent Then, being $\Theta$ $\Lambda$-independent, one has
\barray
\frac{\partial <\Theta >}{\partial \Lambda}&=&
-\int D\Phi\; e^{-S_{eff}}\{Q,\frac{\partial V}{\partial \Lambda} \}
\Theta  \nonumber\\
&=&- <\{Q,\frac{\partial V}{\partial \Lambda} \Theta\}>=0. \label{3.32}
\earray

At this point it is interesting to make contact with the results of Chen
et al. \cite{Chen} on topological features in perturbative CS theory. Using
both dimensional and $trF_{ij}^2$ regularizations, these authors showed
the vanishing of the $\beta$ function for CS theory up to three loops and
also suggested  the absence of diffeomorphism anomaly (which could be present
due to the metric dependence of the $trF_{ij}^2$ regulator). Indeed,
by studying the propagator for the regulated CS model they showed that its
metric dependent transverse part (responsible for the would-be diffeomorphism
anomaly) vanishes because of the unusual properties of the wave-function
renormalization. Our results, summarized by eqs.(\ref{3.22}) and (\ref{3.25})
confirm this suggestion.

Concerning the evaluation of topological invariants using the connection
between CS and TYM theories, let us note that the modifications of the SQ
procedure needed whenever $dim{\cal M}\neq 0$ make the analysis more involved
\cite{Yu}. However, the conclusions of Ref.\cite{Yu} about the connection
between topological invariants of TYM theory and those arising in $d=3$
CS theory should continue to hold when the regulation of the stochastic
process is appropriately taken into account.

We shall conclude this Section discussing
an alternative to the use of  a $\frac{1}{\Lambda^2}
trF_{ij}^2$ regulator, namely, the introduction
of a non-trivial kernel $K_{ij}$
in the Langevin equation in order to have
% for the Chern-Simons systems
a stochastic process which converges to equilibrium.
In fact, this alternative approach was followed
in Refs.\cite{Wu,Fe} in the study of the Abelian and non-Abelian Chern-Simons
models. In this last case, if one chooses $K_{ij}$ in the form:

\beq
K_{ij} =-\epsilon_{ijk}\partial_k, \label{Y1}
\eeq
then the generalized Langevin equation:

\beq
\dot{A}_{i}^{a} = K_{ij}\frac{\delta S}{\delta A_j^a} + \eta_i^a, \label{Y2}
\eeq
yields the correct equilibrium limit {\it to all orders in perturbation
theory} so that conventional propagators are nicely obtained \cite{Wu}.

Now, if starting from eq.(\ref{Y2}) one tries to repeat the steps leading
to the identification of the stochastic partition function associated with
the CS model and the BRST partition function for the TYM theory
(eq.(\ref{3.22})),
there are difficulties originated in the appearence of non-local terms,
related to the presence of kernel $K_{ij}$.

There is however an alternative derivation of the CS-TYM connection presented
in Ref.\cite{Hor1}. As we shall see, it is easy to prove, using this approach,
the  desired connection in the presence of a kernel.
To this end, let us rewrite Langevin equation (\ref{Y2})
and the corresponding Chern-Simons action in the form:

\beq
\dot{A} = K \frac{\delta S_{CS}}{\delta A} + \eta, \label{Y3}
\eeq

\beq
S_{CS} = -\kappa\pi\int_{M_3} d^3x {\cal K}_{0}(x), \label {Y4}
\eeq
where for simplicity we have avoided indices and introduced the notation
${\cal K}_{0}$ for the CS Lagrangian for later convenience
(compare eq.(\ref{2.2})).
The stochastic partition function associated
with the generalized Langevin equation (\ref{Y2}) reads:
\beq
Z_{Stoch}[S_{CS}] = \int D\eta exp[-\int d^3x d^3y dt \;
\eta(x,t) K^{-1}(x,y) \eta(y,t)] \label{Y5}
\eeq
with $K^{-1}(x,y)$ the inverse of kernel $K$ (in fact, one has to slightly
modify $K_{ij}$ in order to make it inversible by adding a longitudinal
part, $K_{ij}\to K_{ij}=-\epsilon_{ijk}\partial_k +
\alpha\partial_i\partial_j$,
and treating $\alpha$ as a gauge fixing parameter).
Now, using eq.(\ref{Y2}) we
can rewrite (\ref{Y5}) in the form:

\begin{eqnarray}
Z_{Stoch}[S_{CS}] &=& \int DA\; det(\frac{\partial}{\partial t} -
K\frac{\delta^2 S_{CS}}{\delta A^2}) \times \label{Y6}   \\
&\times & exp[-tr\int d^3x d^3y dt ( \dot{A} K^{-1} \dot{A} -2\dot{A}
\frac{\delta S_{CS}}{\delta A}  \nonumber\\
&+& K(\frac{\delta S_{CS}}{\delta A})^2 )]. \nonumber
\end{eqnarray}

Consider now the ($3+1$)-dimensional action:

\beq
S^{(3+1)} = tr\int_{M_4} d^3x dt \frac{\delta S_{CS}}{\delta A}\dot{A}
\label{Y7}
\eeq
where gauge fields depend now on an additional (fictitious) time $t$, $t
\in I\equiv [0,T]$ (at the end of
the computations one should let $T \to \infty$).
It is easy to verify that action (\ref{Y7}) is invariant under the following
(large) transformation:

\beq
\delta A = \delta \epsilon(x,t) \label{Y8}
\eeq
provided

\beq
\delta \epsilon(x,0) = \delta \epsilon(x,T) = 0  \label {Y9}
\eeq
Let us use the standard BRST procedure in order to fix this large invariance.
The corresponding partition function reads:

\beq
Z_{BRST}[S^{(3+1)}] = \int DA Db D\chi D\psi exp[-S^{(3+1)} -tr\int_{M_4}
d^3xdt \{Q,\chi(F - \frac{b}{2})\}]
\label {Y10}
\eeq
with $b$ a Lagrange multiplier enforcing the gauge condition (which at the
end will turn to be $F=0$) and $\chi$ and $\psi$ the ghost fields (which
in our previous derivation were introduced through eqs.(\ref{3.13})).
BRST transformations $\{Q,\quad\}$ are defined as:

\beq
\begin{array}{lll}
\{Q,A\} = \psi  & &  \{Q,b\} = 0 \nonumber \\
\{Q,\chi\} = K^{-1}b  & & \{Q,\psi\} = 0   \label{Y11}
\end{array}
\eeq
With this, we have:

\barray
Z_{BRST}[S^{(3+1)}] &=& tr\int DA Db D\chi D\psi \times \label{Y13} \\
&\times &exp[-S^{(3+1)} -\int_{M_4} d^3xdt
(K^{-1} b (F - \frac{b}{2}) + \chi \frac{\delta F}{\delta A} \psi)]
\nonumber
\earray
or, integrating out $b$:

\beq
Z_{BRST}[S^{(3+1)}] = \int DA D\chi D\psi exp[-S^{(3+1)}
-tr\int_{M_4} d^3xdt
(\chi \frac{\delta F}{\delta A} \psi +\frac{1}{2} K^{-1}F^2)] \label{Y14}
\eeq
Then, if we choose as gauge function

\beq
F = \dot{A} - K \frac{\delta S_{CS}}{\delta A} \label{Y15}
\eeq
we get:

\begin{eqnarray}
Z_{BRST}[S^{(3+1)}] = \int DA det(\frac{\partial}{\partial t} - K\frac{\delta^2
S_{CS}}{\delta A^2}) & \times & \nonumber\\
\times exp[-tr\int_{M_4} d^3x dt (\dot{A} K^{-1} \dot{A}
- 2\dot{A}\frac{\delta S_{CS}}{\delta A}&+&
K(\frac{\delta S_{CS}}{\delta A})^2 ) \label{Y16}
\end{eqnarray}
and comparing with eq.(\ref{Y6}) we have:

\beq
Z_{BRST}[S^{(3+1)}] =Z_{Stoch}[S_{CS}]
\label{Y17}
\eeq

It remains to show that $S^{(3+1)}$ is indeed the classical action
associated with TYM theory. To this end note that:

\beq
S^{(3+1)} = Q[T] - Q[0] \label{Y18}
\eeq
where:

\beq
Q(t) \equiv -\kappa\pi\int_{M_3} d^3x {\cal K}_{0} \label{Y19}
\eeq
Now, if we define a gauge field $A_{\mu}$ in $(3+1)$-dimensions (in
the $A_0 = 0$ gauge) as $A_{\mu} =(A_i,0)$, one can easily prove \cite{JL}
that:

\beq
\lim_{T \to \infty} Q[T] =
-\kappa\pi\int_{M_4} d^4x \partial_{\mu}{\cal K}_{\mu} \label{Y20}
\eeq
where ${\cal K}_{\mu}$ is the Chern-Simons characteristic class:

\beq
{\cal K}_{\mu} = -\frac{1}{16\pi}tr \epsilon_{\mu \nu \alpha \beta}
(F_{\nu \alpha}A_{\beta}
-\frac{2}{3} A_{\nu} A_{\alpha} A_{\beta}) \label{Y21}
\eeq
(One has to choose initial conditions $A_i(x,0)=0$ so that $Q[0]=0$). One then
has, after taking the $T \to \infty$ limit:

\beq
S^{(3+1)} = -\kappa\pi\int_{M_4} d^4x \partial_{\mu}{\cal K}_{\mu} \label{Y22}
\eeq
or

\beq
S^{(3+1)} = \frac{\kappa}{16\pi}tr \int_{M_4} d^4x\; {^*}\! F_{\mu\nu}
F_{\mu\nu} \label{Y23}
\eeq

That is, action $S^{(3+1)}$ coincides with the Chern-Pontryagin invariant
which was taken by Baulieu and Singer \cite {BaS} as starting classical
action to obtain, after BRST quantization, Witten's
TYM theory. Then, eq.(\ref{Y17}) states the equivalence
between the stochastic partition function for CS theory with kernel $K$
and the partition function for TYM theory (constructed after quantization
of classical action (\ref{Y23})).

A comment is here in order: we have not introduced in our derivation
neither ordinary ghosts associated with ordinary gauge invariance nor
ghost for ghost, necessary to fix the second generation gauge-invariance
characteristic of TYM theory.
We have ignored them just to present more clearly our arguments, but these
ghosts can be easily included following the same steps leading from
eq.(\ref{3.8}) to eq.(\ref{3.17}) and our proof remains still valid.
In any case, we have
shown that the equivalence between TYM and CS partition functions can be
also proven when a non-trivial kernel is introduced in the Langevin equation
in order to handle the convergence to equilibrium problem which arises for
CS theory. The interesting fact is that in this approach convergence can
be proven to all orders in perturbation theory \cite{Wu}.

\section{Conclusions}

In the last few years the interest in Topological Field Theories has
prompted cross-fertilizing investigations both in Mathematics and in Physics.
One important result of these studies is that they allow to
establish multiple connections between field-theoretical models in different
number of space-time dimensions.

One example of this is the connection
discovered by Witten \cite{Wi2} between $d$=3 Chern-Simons gauge theory and
$d$=2 rational conformal field theories: the states in the Hilbert space
of the CS theory defined on a compact surface is the space of conformal
blocks of rational conformal field theories.
Another example is the relation between $d$=3 CS theory and $d$=4 Topological
Yang-Mills theory \cite{Yu,Bau2}: topological invariants in TYM
theory correspond to observables in CS theory.

These connections have been established using different
approaches:
path-integral quantization, canonical quantization, stochastic
quantization. This last approach is the one we have employed
in the present work to re-analyze the relation between
$d$=3 CS and $d$=4 TYM theories (although other relations like
for example that connecting $d$=1 spin model and $d$=2
topological sigma model \cite{FMS} can be studied using an
identical approach).

The original studies \cite{Yu}-\cite{CLS} of the CS-TYM connection
using SQ were somehow
formal in the sense that the non-con\-vergence of the underlying stochastic
process was disregarded. Later on, the problem of convergence in SQ of CS
theory was solved \cite{Wu,Fe} by introducing appropriate kernels or
regulators so that a correct equilibrium limit can be attained. After
these results, it remained to determine whether the connection between
CS and TYM theories was still valid in the presence of regulators.
One of the aims of the present work was to address to this question
and the answer is affirmative.
As we have shown in Section 3, the addition of a regulator
$\frac{1}{\Lambda^2}trF_{ij}^2$ to the CS action does not affect the
proof of equivalence between the stochastic partition function for CS
theory and the BRST partition function for TYM theory.

Indeed, due to the fact that the effective 4-dimensional action $S_{eff}$
is a $Q$-exact form even in the presence of a regulator $\Lambda$
(see eq.(\ref{3.19})), the resulting partition
function $Z^{TYM}_{BRST}[\Lambda]$
is independent of $\Lambda$. Hence, it can be evaluated in the limit
$\Lambda \to \infty$ where it coincides with the TYM partition function
defined by Witten \cite{Wi1}. In any case, for arbitrary $\Lambda$,
$Z_{BRST}^{TYM} [\Lambda]$ defines a topological theory, since the
corresponding v.e.v. for the energy-momentum tensor vanishes (see
eq.(\ref{3.27})) and all topologically invariant observables are
$\Lambda$-independent (see eq.(\ref{3.29})).

In the course of this work we have also found of interest to study,
in the same vein
as Namiki et al. \cite{Na} did for the SQ of Yang-Mills theory, how convergence
for non-Abelian CS model is attained when the regulator is introduced and
the precise way in which standard propagators are obtained using SQ.
We have proven that the addition of a regulator
$\frac{1}{\Lambda^2}trF_{ij}^2$ to the CS action renders the stochastic
process associated with the CS Langevin equation convergent. The proof we
have presented is valid up to second order in perturbation theory and states
that the transverse part of the stochastic diagrams reproduces the standard
gluonic CS propagator (see eq.(\ref{2.24})). It is interesting to note that
the way in wich stochastic diagrams for CS regulated theory sum up to the
standard propagator is basically identical to the way it works in the case
of pure YM theory.

We have also discussed the alternative regularization scheme which consists
in the introduction of a non-trivial kernel. Indeed, an appropriate choice
of the kernel  \cite{Wu,Fe} (see eq.(\ref{Y1})) yields the correct equilibrium
limit {\it to all orders in perturbation theory}. As we have shown, also
in this approach it becomes clear that the equivalence between the stochastic
partition function for CS theory and the BRST partition function for TYM
theory can be proved in the presence of a regulator (in this case a kernel).

\newpage
TABLE CAPTIONS.
\vspace{1cm}

Table 1: Diagrams for stochastic vertices.

\vspace{2cm}

FIGURE CAPTIONS.
\vspace{1cm}

Figure 1: Perturbative expansion of $A_i^a(k,t)$.

Figure 2: Graphical representation of ${\cal D}_{ij}^{ab}(k,t)$.

\end{document}